# How a SnakeCanRaise its Neck High although its Body is Very Soft. Similarity with a Thin Plastic Tube Filled with Air

Handika Dany Rahmayanti, DesyanaOlenka Margaretta, Nadya Amalia, Fisca Dian Utami, ElfiYuliza, Rahmawati Munir, Nova LailatulRizkiyah, and Mikrajuddin Abdullah[a]

Department of Physics
Bandung Institute of Technology
JalanGanesa 10, Bandung 40132, Indonesia
[a]E-mail: mikrajuddin@gmail.com

**Abstract**

The bodies of snakes, such as king cobras, are very soft so that it is nearly impossible for them to raise their headshigher than around one meter. By direct measurement we were only possible to hold the body of a freshly killed cobra vertically at around 0.05 m. Here, for the first time it is reported that the king cobra can control the effective elastic modulus of its body so that it can raise its head up to several tens of centimeters. The effective elastic modulus is enhanced by increasing the pressure of air trapped inside the respiration channel, which is similar to increasing the stiffness of a thin plastic tube by filling it with air atpressure above atmospheric pressure. The neck height increases with the effective elastic modulus according to a scaling relationship. It was also simply proved that the peak of force or constriction pressure is proportional to the snake's diameter. This work may provide a physical foundation underlying the mechanical properties of slender animals.

## INTRODUCTION

In relaxed state (resting or crawling) the body of the king cobra (*Ophiophagushannah*) can be compared to a circular slender rod [1]. The body's tissue is very soft so it is difficult for the cobra to raise its neck high. To prove this, the body of a freshly killed cobra was held in a vertical position. The cobra was obtained from a restaurant in Bandung city, Indonesia, where foods from cobra meat are served. It was found that the body can only be held vertically at less than 0.05 m, above which it bends down (**Fig. 1**(a)).

Therefore, the question is: how can a king cobra and other snakes raise their head several tens of centimeters, as commonly observed?

The standard body length of a king cobra is between 3 and 4.5 m [2,3]. It can raise its neck about one third of its body length, i.e. up to 1-1.5 m, which means that the theoretical result is very small. The maximum height occurs when the snake is ready to peck. We also measured directly how high the cobra can raise its neck in a cobra attraction place in Kuningan regency, Indonesia with the help of a cobra handler and found an estimated height of 0.5 m.

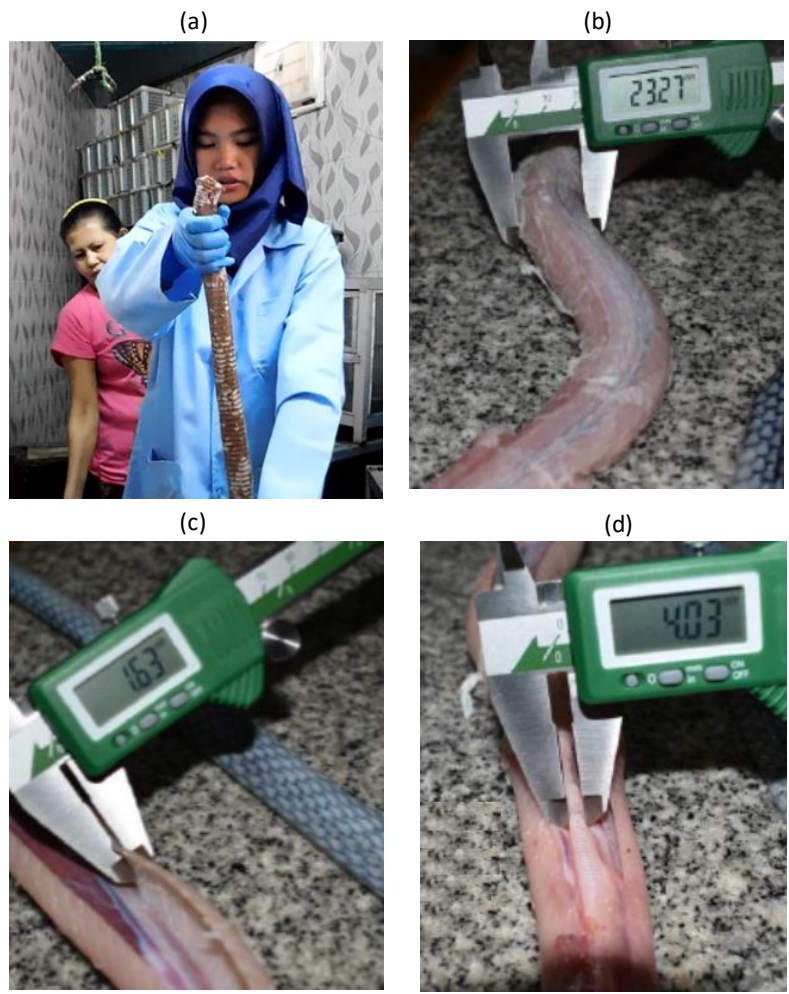

**Figure 1** Measurement of cobra body dimension: (a) measuring how high a freshly killed cobra body without head can be held vertically, (b) outer diameter of a freshly killed cobra, (c) thickness of the cobra muscle along the neck, and (d) diameter of digestive or respiration channel.

Let us first explain this phenomenon from the concept of self-buckling of a thin cylinder. The critical height for self-buckling is given by a classical formula [4]:

$$H_{max} = \frac{Y}{\rho g \sqrt{3(1-v^2)}} \left(\frac{t}{R}\right) \quad (1)$$

where $v$ is the Poisson ratio, $R$ is the cylinder radius, $t$ is the wall thickness, $\rho$ is the wall density, and $g$ is the acceleration of gravitation. The Poisson ratio of muscle is around 0.5 [5]. The average mass density of cobra muscle was measured directly and found to be 1.670 kg/m$^3$. The cross-section of the cobra body was also measured and it was found that the outer diameter (excluding skin) was $d_o$ = 0.02327 m (**Fig. 1(b)**) and the body cross-section was like a hollow cylinder with a wall thickness of around 0.00163 m (**Fig. 1**(c)), or an inner diameter of $d_i$ = 0.02001 m. This is the typical cross section of a snake's body [6]. Along the neck, a digestive or respiration channel with a diameter of around 0.00403 m was observed (**Fig. 1(d)**). The body muscle is covered by very thin skin of thickness around 0.00025 m. Based on this geometry it is arguable to consider the cobra body around the neck as a thin cylinder.

The elastic modulus of cobra muscle was estimated to be comparable to that of humans or other animals. For example, the tension and compression elastic moduli of human muscle are 10-15 kPa and ≈ 25 kPa, respectively [7]. The elastic modulus of human muscle measured using supersonic shear imaging was obtained between 5.6 and 18.3 kPa [8]. Using magnetic resonance elastography, the elastic modulus of human muscle was obtained at 12.3 ±0.5 kPa[9], 27 kPa[10], and 17.9±5.5 kPa[11]. The elastic modulus of soft tissue from bovine muscle measured using an ultrasound method was between 1.46 kPa – 3.15 kPa, while using the instron method 1.2-1.8 kPa was found [12].

Let us assume that the elastic modulus of cobra muscle is comparable to that of the upper bound of human muscle, i.e. around 25 kPa. Using Eq. (1) the critical height for self-buckling was estimated at around 0.14 m. This figure becomes smaller when the middle values of the measured elastic moduli of the human muscle are used. For example, compared with the middle value of about 15 kPa, the critical height for self-buckling was estimated at around 0.085 m. This value is close to our direct measurement result of the freshly killed cobra of around 0.05 m, as shown in Fig. 1(a).

Other snakes that can also raise their heads, although not as high as the king cobra, are the cotton mouth snake (*Agkistrodonpiscivorus*) [13], the eastern brown snake (*Pseudonajatextilis*) [14], and the southern banded water snake (*Nerodiafasciata*) [14]. There must be a specific mechanism that controls the raise of the cobra's head. The purpose of this work was to investigate this very challenging mechanism. Understanding the mechanism of a living thing may open opportunities for developing new technologies by biomimicry or bioinspiration.

**METHOD**

**Thin Plastic Tube Demonstration**

Based on the direct measurement explained above, the snake or cobra body is like a thin and soft tube. In normal situations it is nearly impossible to hold the tube vertically, but cobras can. To investigate the reason, a simple experiment was done using a specific material that mimics the cobra body, i.e. a plastic tube made of polyethylene terephthalate (PET), as shown in **Fig. 2 (a)**. This comparison is arguable since the tube shape resembles the respiratory system of a snake. Snakes have a small opening just behind the tongue named glottis, which opens into the trachea (windpipe) [15]. The trachea is a long, strawlike structure supported by cartilaginous rings and terminates right in front of the heart (**Fig. 1(d)**). Around 25% of the snake's body length measured from the head consists of the head, the esophagus and trachea, and the heart. This section is most likely a hollow tube.

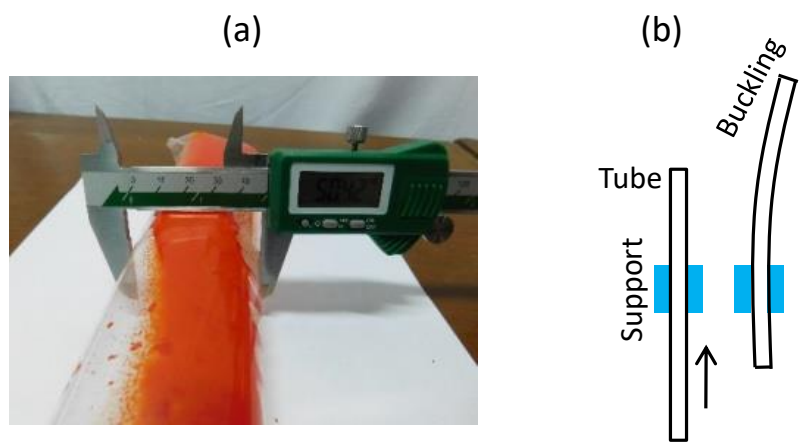

**Figure 2.** (a) The plastic tube used in the experiment and (b) the shape of the plastic tube when progressively pushed up between vertical supports: (left) prior to buckling and (right) buckling state.

The thickness of the tube sheet was 0.03 mm. Tube diameter $d$ was varied at (a) 0.02521 m (b) 0.021485 m (c) 0.01844 m (d) 0.01571 m, and (e) 0.01242 m. The tube was filled with air using a hand pump of chamber length $L = 0.16$ m and diameter $D = 0.05$ m. It was assumed that the ideal gas law was applicable to estimate the pressure inside the tube. If the tube length is $\ell$, the volume of space inside the tube is $\pi d^2 \ell/4$. The volume of air in the pump cavity is $\pi D^2 L/4$. The pressure of the air in the pump space is equal to the atmospheric pressure $P_0$. One end of the tube was closed tightly and the other end was connected with the pump. This is similar to the snake glottis, which is always closed, forming a vertical slit unless the snake takes a breath[15]. If the pump is pressed $n$ times, the pressure inside the plastic tube space, $P$, satisfies $P(\pi d^2 \ell/4) = nP_0(\pi D^2 L/4)$, or

$$P = \frac{D^2 L P_0}{d^2 \ell} n \tag{2}$$

Pumping air into the tube is comparable to the inspiration process in the snake. Inspiration is an active process (muscles contract), whereas expiration is passive (muscles relax).

After filling the tube with air, it was held vertically. The upper end of the tube was placed between vertical supports and the tube was gently pushed up vertically. Initially, the tube could stay vertical. After that it suddenly bent after reaching a certain height (**Fig. 2(b)**). This condition is assumed to be the self-buckling condition. The length of the tube when the buckling occurs is known as the critical height for self-buckling.

**Simulation**

Let us divide the cobra neck of length $L$ into $N$ identical segments, $a$ ($= L/N$). The segment at the free end is the 1$^{st}$ segment and the one at the clamp (fixed end) is the $N$-th segment. The $j$-th segment has angle $\theta_j$ ($j = 1$ to $N$) to the horizontal. When a certain load, $W_1$, is exercised on the 1$^{st}$ segment, the bending angle of the $j$-th segment satisfies [16]:

$$\theta_j = \theta_{j-1} - \frac{a^2 W_1}{Y_j I_j} \sum_{k=1}^{j-1} \cos\theta_k - \frac{a^3 g}{Y_j I_j} \sum_{i=1}^{j-1} \sum_{k=1}^{i} \lambda_k \cos\theta_i \tag{3}$$

where $Y_i$, $I_i$, and $\lambda_i$ are the elastic modulus, area moment, and mass per unit length of the $j$-th segment, respectively, and $g$ is the acceleration of gravitation. The angles of all segments were calculated by selecting the angle of the first segment until the fixed (*N*-th) segment satisfied the boundary condition.

For measuring the coordinates of the neck profile, two perpendicular meter sticks were used as length references. The images were recorded at the aforementioned cobra attraction place in Kuningan regency. The images were tracked using the Tracker software application, an Open Source Physics (OSP) Java framework-based image and video processing program[17], to predict the neck thickness from the head to the neck base and the coordinates of the neck curve. The use of recorded images or videos followed by tracking for investigating the mechanical properties of snakes is not new [18,19]. From the measurements, the change in cross section with position was obtained as well as the bending angle, which was used for estimating the elastic modulus.

When the cobra raises its head, the neck thickness is inhomogeneous and the cross section is not completely circular (**Fig. 3(a)**). At the neck base, the cross section can be approximated by a circle but moving to the head, the cross section deviates from a circular shape. Thus, the simulation must consider this variation in geometry.

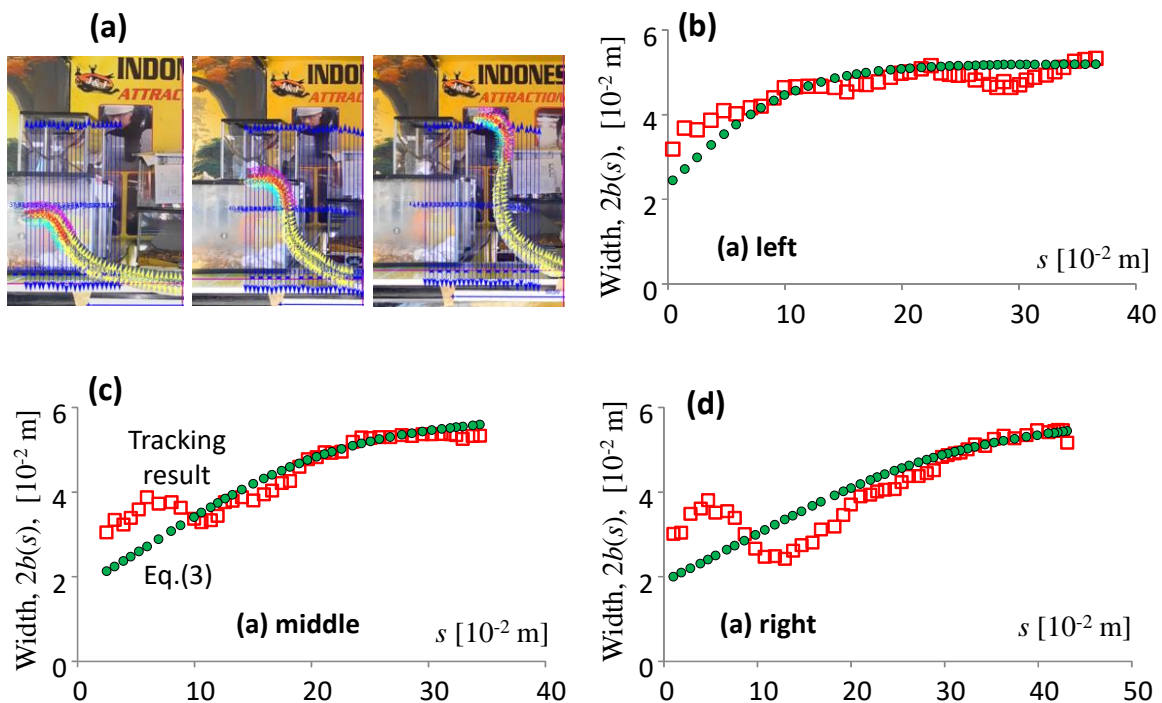

**Figure 3** (a) Measurement of cobra neck coordinates and thickness at the *cobra attraction place* in three positions. (b)-(d) The corresponding measurement results of cobra neck thickness as a function of position (open squares) at three different positions and the calculated results of thickness using Eq. (4): (b) first position ($A = 0.026$ m, $\xi L/2 = 0.01$ m, $\gamma = 10$ m$^{-1}$), (c) second position ($A = 0.029$ m, $\xi L/2 = 0.07$ m, $\gamma = 6$ m$^{-1}$), and (d) third position ($A = 0.029$ m, $\xi L/2 = 0.09$ m, $\gamma = 4$ m$^{-1}$).

For calculation purposes, the cobra neck was modeled in emergency condition to have a circular cross section at the bottom and an elliptical cross section toward the head. The semi-minor axis, $b$, is assumed to decrease when moving toward the head. The true dependence of $b$ on the distance along the neck is unknown. The tracking results are indicated with open squares in **Figs. 3(b)-(d)**. Three images of the neck with different lengths were tracked. It was identified that the data fairly fit the following function:

$$2b(s) = A\left[1 + \tanh[\gamma(s - \xi L/2)]\right] \qquad (4)$$

with $\xi$ is a constant, $0<\xi<1$ and $0 <s<\xi L$. The square symbols in **Figs. 3(b)-(d)** are the calculated results using Eq. (4). The fitting equation can explain very well the thickness at positions far from the head, but a slight deviation was observed around the head. This may be due to inaccuracy when recording the image, or inaccuracy in the tracking process. The function in Eq. (4) was selected for several reasons. The function must yield asymptotic values at positions far away from the neck center in opposite directions. This is easily satisfied by Eq. (4).

The fitting results using Eq. (4) are shown by closed squares in **Figs. 3(b)-(d)**. The parameters of fitting were: (A) ($A = 0.026$ m, $\xi L/2 = 0.01$ m, $\gamma = 10$ m$^{-1}$), (B) ($A = 0.029$ m, $\xi L/2 = 0.07$ m, $\gamma = 6$ m$^{-1}$), and (C) ($A = 0.029$ m, $\xi L/2 = 0.09$ m, $\gamma = 4$ m$^{-1}$). The closeness of the approximation was also calculated by defining parameter $\varepsilon = \sqrt{(1/N)\sum_{i=1}^{N}(w_{fit,i} - w_{meas,i})^2}$ with $N$ is the number of measured data, $w_{meas,i}$ is the $i$-th data of width from measurement, and $w_{fit,i}$ is the $i$-th data of the width from fitting. We obtained (A) $\varepsilon = 0.0034$ m, (B) $\varepsilon = 0.0045$ m, and (C) $\varepsilon = 0.0059$ m. These results demonstrate that Eq. (4) is acceptable as fitting equation.

When changing the neck shape from circular to elliptical, the most probable conserved quantity is the body's circumference. The elliptical circumference is well approximated by $\pi\sqrt{2}\sqrt{a^2+b^2}$ [20] with $a$ is the semi-major, and equal to $2\pi r_0$ so that we have the following equation: $a = \sqrt{2r_0^2 - b^2}$. Since the cross-section depends on position, the area moment also depends on position. For a hollow circular cross-section with inner and outer radii $r_i$ and $r_o$, respectively, we have $I_c = (\pi/4)(r_0^4 - r_i^4) = (\pi/4)(r_0^4 - (r_0 - \delta)^4)$, with $\delta$ is the muscle thickness. For a hollow elliptical cross section, the area moment is $I_e = (\pi/4)\left(b^3 a - (b-\delta)^3(a-\delta)\right)$.

## RESULTS AND DISCUSSION

**Plastic Tube**

**Figure 4** is the plot of critical height for self-buckling against the pressure inside a tube for different diameters: (circle) 0.01242 m, (diamond) 0.01571 m, (square) 0.01844 m, (triangle) 0.021485 m, and (star) 0.02521 m. It can clearly be seen than when the pressure inside the tube wassmaller than $P_0 \approx 1$ atmit was impossible to hold the tube vertically. This is because the air pressure inside the tube wassmaller than the pressure outside, leading it to crumple. The tube can be held vertically if the pressure inside is higher than atmospheric pressure. This strongly suggests that a very weak tube can be held vertically by increasing the air pressure inside the tube cavity. The critical height increases when increasing the pressure of the air inside the cavity as long as the pressure surpasses the atmospheric pressure. It can also be seen from the figure that the critical height for buckling is weakly sensitive to the tube diameter. The data for all tube diameters nearly coincide. Therefore, the main factor controlling the critical height for buckling is the pressure instead of tube diameter. Next, we will inspect how the critical height for buckling changes with excess pressure $P-P_0$ for different tube diameters.

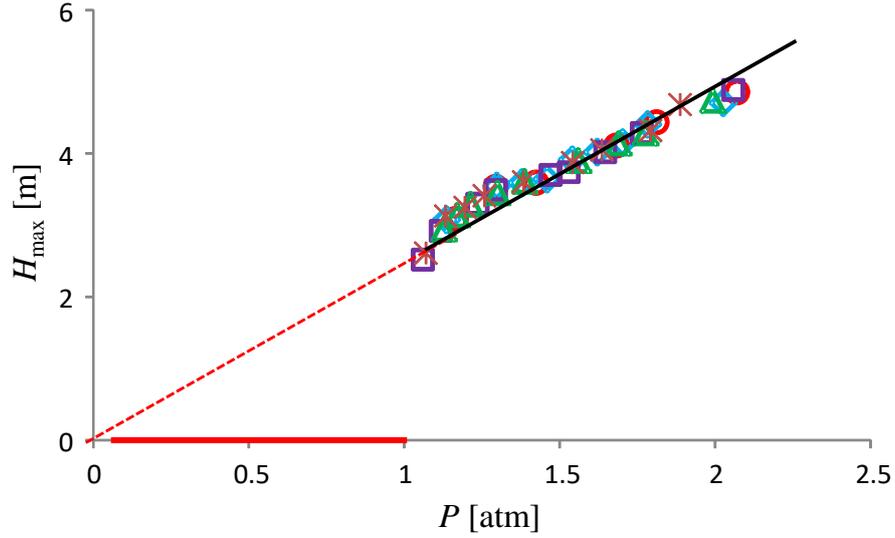

**Figure 4** Effect of air pressure inside a tube on critical height for buckling. Five tube diameters were used: (circle) 0.01242 m, (diamond) 0.01571 m, (square) 0.01844 m, (triangle) 0.021485 m, and (star) 0.02521 m.

The difference in air pressure inside and outside of the liquid layer is controlled by the surface tension, γ, according to the relation:

$$\Delta P = P - P_0 = \gamma \left( \frac{1}{R_1} + \frac{1}{R_2} \right) \qquad (5)$$

with $R_1$ and $R_2$ are the curvatures in perpendicular direction. For a tube with a cylindrical shape we have $R_1 = R$ (tube radius) and $R_2 \to \infty$ so that $P - P_0 = \gamma / R$, or

$$\gamma = (P - P_0) R \qquad (6)$$

From Eq. (6) we proposethat the surface tension of the tube wall increases with the amount of excess pressure. In other words, the surface tension is not a constant parameter but depends on the amount of excess pressure. The tube becomes stronger when the amount of excess pressure increases. Therefore, the tube becomes stronger when the surface tension increases. Based on this argument, we can hypothesize that the *effective* elastic modulus of the tube filled with air is proportional to the surface tension, $Y \propto \gamma$. This is a crucial hypothesis that may need further justification. Since only Y and R are variables, we can write $H_{max} \propto Y/R$. Using Eq. (6) and remembering assumption $Y \propto \gamma$ we finally find:

$$H_{max} \approx (P - P_0) \qquad (7)$$

It is very clear from Eq. (7) that $H_{max}$ depends only on the air pressure and is unaffected by the tube's diameter, as confirmed by **Fig. 4**. It is also clear that $H_{max}$ increases linearly with pressure, as also confirmed by **Fig. 4**.

If both ends of the plastic tube are closed and the small volume inside is filled with air, the plastic takes on the shape of a slender circular rod. The structure is soft and it is very difficult to hold it vertically. However, when the air volume inside the tube is increased, the pressure increases. The tube can be held vertically when the pressure of the air inside surpasses the critical pressure. Above the critical pressure, the tube can be held vertically until a specific height, above which buckling occurs. The critical height for buckling increases with the increase of the pressure inside the tube.

In the present work it was shown that cobras can raise their necks several tens of centimeters by controlling the elastic modulus of their neck. It was also shown that the elastic modulus of the cobra's neck has a scaling relationship with neck length. For this purpose, the continuum equation for describing bending of slender rods/beams was first transformed into a discrete form. The governed equation can be applied to slender beams in any situation: large bending, inhomogeneous beams, inhomogeneous cross-sections, etc.

**Simulation Results**

First, the condition when a cobra makes a free end angle of 0° to the horizontal and the fixed end makes an angle of -32° was simulated. Based on Eq. (3), information is needed on the mass of the neck per unit length. This quantity was estimated based on the measurement of the size of the cobra at the cobra attraction place, where the body diameter in relaxed condition was around 0.05 m. Therefore, the circumference of the cobra body was $\approx \pi d$ = 0.157 m. The muscle thickness was taken as nearly the same as that measured at the restaurant of 0.00163 m so that the cross section was approximately 0.000256 m². To get accurate results, the density of the freshly killed cobra meat purchased from the restaurant mentioned previously was measured, obtaining 1,670 kg/m³ so that the mass per unit length was around λ = 0.428 kg/m. Of course this value is very small when compared to the average measures of cobras, where the common length is 3.2 meter and mass 4.9 kg [21] so that the mass per unit length is approximately 1.53 kg/m. The last value is mainly contributed by the cobra's digestive system. In the simulation, only considered the neck system was considered, which is like a thin hollow pipe.

**Figure 5** shows examples of the simulation results. The neck length was varied at (a) 0.16 m, (b) 0.24 m, (c) 0.32 m, and (d) 0.48 m to find the elastic moduli that match these boundary conditions. The portions of the neck that have an elliptical cross section were varied at (A) 25% of the neck length, (B) 50% of the neck length, and (C) 70% of the neck length.

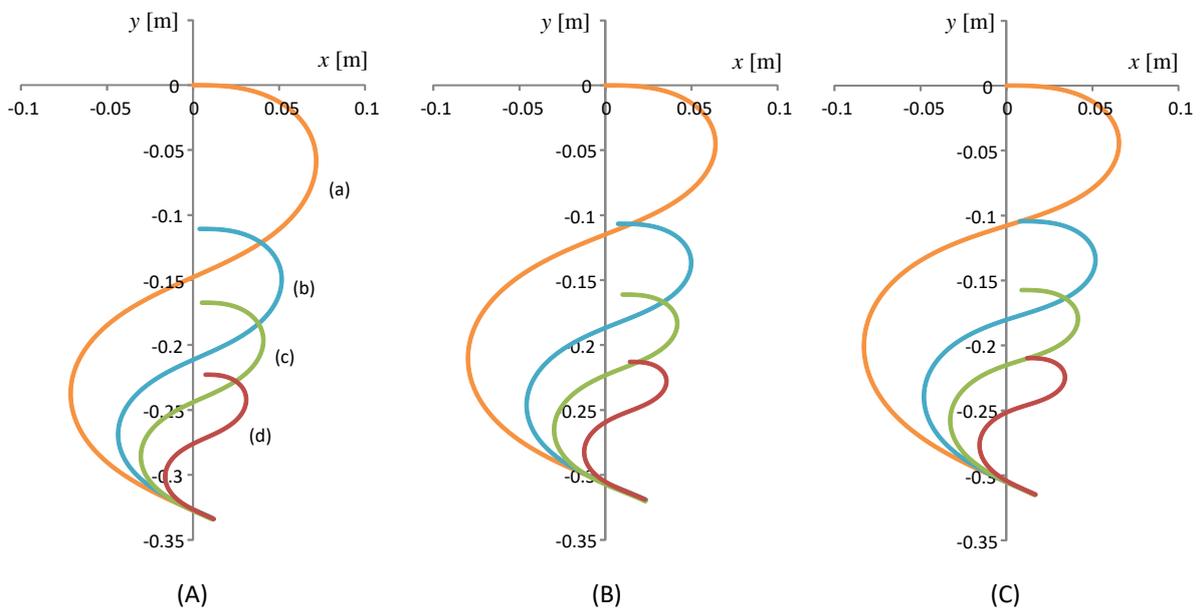

**Figure 5**. Simulation results of the cobra neck shape at different lengths: (a) neck length = 0.16 m, (c) 0.24 m, (c) 0.32 m, (d) 0.48 m and different portions of the neck with an elliptical cross section: (A) 25% of neck length, (B) 50% of neck length, and (C) 70% of neck length.

It is clear from **Fig. 5** that the shape obtained from the simulation approached the shape of the cobra neck inemergency condition, like a mirrored letter S. Different shapes were calculated using different elastic moduli as summarized in **Table 1**.

Let us inspect how the elastic modulus changes with neck length at different portions of the elliptical cross section. **Figure 6(A)** is the log-log plot of the elastic modulus as function of neck length. It can be seen that all curves have exactly the same slope (= 2) with $R^2 \approx 1$ for all portions of the elliptical cross section. Therefore, the elastic modulus changes with neck length according to

$$Y \propto L^2 \tag{8}$$

This result is exactly the same as the equation that describes the critical load for buckling, $P_c = \pi^2 YI/(KL)^2$. In the simulation, the load originated from the cobra's head is such that it is unchanged for all neck lengths to maintain $Y \propto L^2$, which is consistent with **Figure 6(A)**.

**Table 1** Simulated elastic moduli of cobra neck at different neck lengths and different portions of the neck with an elliptical cross section.

| Neck length [m] | Portion of the elliptical cross section [%] | Elastic modulus, $Y$ [$\times 10^8$ Pa] |
|---|---|---|
| 0.16 | 25 | 0.48 |
|  | 50 | 0.57 |
|  | 70 | 0.67 |
| 0.24 | 25 | 1.09 |
|  | 50 | 1.28 |
|  | 70 | 1.5 |
| 0.32 | 25 | 1.95 |
|  | 50 | 2.3 |
|  | 70 | 2.7 |
| 0.48 | 25 | 4.4 |
|  | 50 | 5.2 |
|  | 70 | 6 |

The ability of snakes to control the strength of their muscles is commonly known. An example of this mechanism is constriction by a snake, which immobilizes and kills prey by using body loops to exert pressure [22]. Constriction can disrupt breathing [23,24] and circulation [23,25] of the prey.

It is interesting to see whether the elastic modulus of the cobra neck depends of the angle of the free end (the end at which the head is located). Two additional angles for the free end were simulated, i.e. -30° and +30° for different neck lengths and different portions of the elliptical cross section. We conclude that as the free end angle increases (from negative to positive), the neck profile becomes shorter. For example, for a neck with a length of 0.48 m and portion of the elliptical cross section of 50%, the vertical distance of the neck when the cobra makes a free end angle of -30° is around 0.4 m, becomes 0.32 m when the cobra makes a free angle of +0°, and becomes 0.225 m when the cobra makes a free angle of +30°. It also implies that the neck profile becomes wider when the free end angle increases.

The cobra controls the neck shape by changing its elastic modulus. To make the free end angle more negative, the cobra must increase the elastic modulus. For example, for a neck of length 0.48 m and the portion of the elliptical cross section of 50%, the cobra must produce an elastic modulus of $5.8 \times 10^8$ Pa to make an angle of -30°, $5.2 \times 10^8$ Pa to make an angle of -30°, and $4.75 \times 10^8$ Pa to make an angle of +30°. The same change is also observed for different neck lengths and different portions of the elliptical cross section.

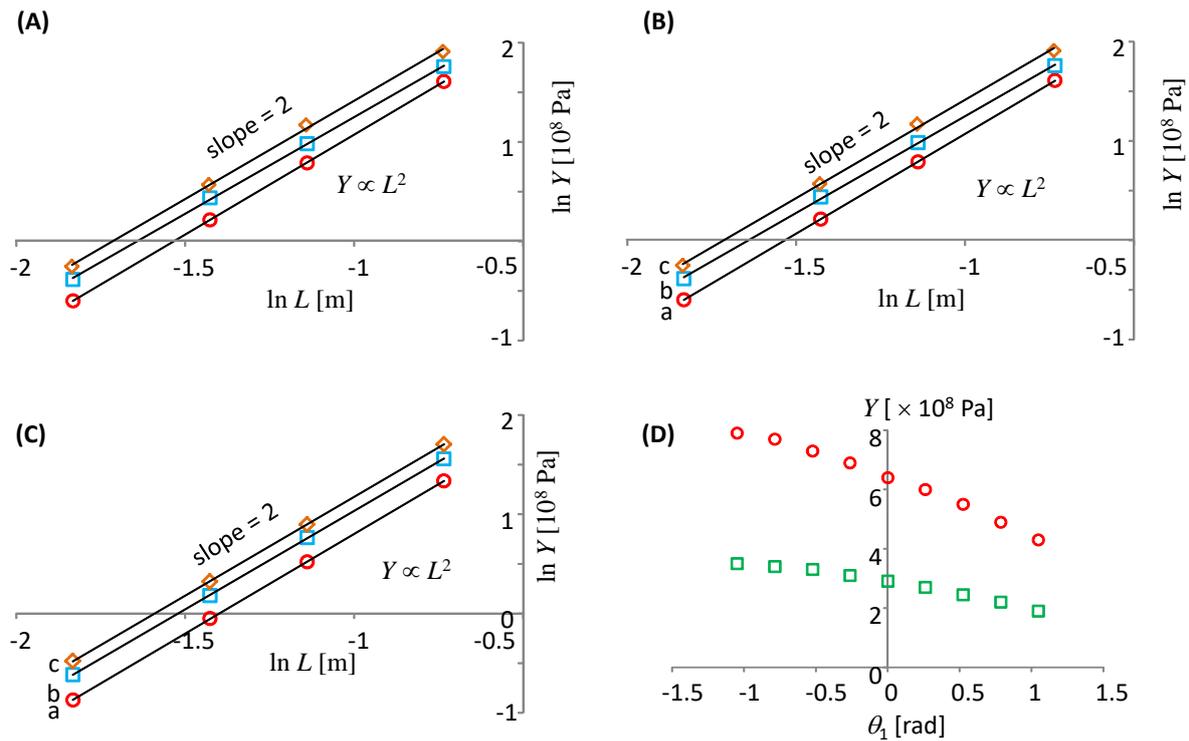

**Figure 6** Dependence of ln $Y$ with respect to ln $L$ for different portions of the neck with an elliptical cross section: (a) 25% of the neck length, (b) 50% of the neck length, and (c) 70% of the neck length. (A) is for the free end making an angle of 0°, (B) is for the free end making an angle of -30°, and bottom figure (C) is for the free end making an angle of +30°.

(D) The elastic modulus as a function of free end angle for different neck lengths (upper) 0.48 m and 0.32 m (bottom). The portion of the elliptical cross section was fixed at 50%.

**Figures 6(B)-(C)** are the log-log plot of the elastic modulus as a function of neck length at different free end angles: (B) -30° and (C) +30°. It can be seen that all curves from both figures have exactly the same slope (= 2) with $R^2$ very close to unity for all portions of the elliptic cross sections. Therefore, it was identified that the elastic modulus changes with neck length according to Eq. (8).

**Figure 6(D)** shows the change of elastic modulus as a function of the free end angle. Two neck lengths were simulated: 0.48 m and 0.32 m and the portion of the elliptical cross section was fixed at 50%. It is clear that the elastic modulus monotonically decreases with the increase of the free end angle. The change is not perfectly linear but slightly curves downward.

For a slender rod with uniform mass density per unit length, Eq. (3) can be rewritten as $Y_j I_j (\theta_j - \theta_{j-1})/a = -aW_1 \sum_{k=1}^{j-1} \cos\theta_k - a^2 g\lambda \sum_{i=1}^{j-1} i\cos\theta_i$. If the segment length is very small, then $(\theta_j - \theta_{j-1})/a = (d\theta/ds)_i$. Therefore, the right-hand side of this equation is the z-th component of the flexion moment and we can write $M_z = -aW_1 \sum_{k=1}^{j-1} \cos\theta_k - a^2 g\lambda \sum_{i=1}^{j-1} i\cos\theta_i$. If the outer diameter of the cobra body is $r$ and the inner diameter is $r-\delta r$ with $\delta r \ll r$, then $\lambda \propto \pi [r^2 - (r-\delta r)^2] \propto 2\pi r \delta r$. Assume the thickness of the head wall as $\Delta r$, which also satisfies $\Delta r \ll r$, then $W_1 \propto \pi [r^2 - (r-\Delta r)^2] \propto 2\pi r \Delta r$. Therefore, the dependence of $M_z$ on snake body diameter is approximated by $M_z \propto r$. Since the flexion moment is generated by force acting on the cobra body, it can be said that the force produced by the cobra body satisfies $F \propto r$. This result implies that the peak of force or constriction pressure is proportional to the snake's diameter, exactly the same as recently reported by Penning and Moon [6].

CONCLUSION

It was demonstrated for the first time that king cobras can control the effective elastic modulus of their neck muscle when raising the head/neck. It was hypothesized that such enhancement is facilitated by increasing the air pressure in the cavity of the respiration channel. The elastic modulus increases with neck length. It was also shown that the elastic modulus square scales with neck length. Finally, it was simply shown that the peak of force or constriction pressure is proportional to the snake's diameter as reported by Penning and Moon.